\documentclass[12pt]{article}
 \pdfoutput=1
\usepackage[table]{xcolor}
\usepackage{amsmath,amssymb,exscale}
\usepackage{graphicx}
\usepackage{epsfig}
\usepackage{multicol}
\usepackage{color}
\usepackage{mathrsfs}
\usepackage{blindtext}
 \usepackage{fancyhdr}
\usepackage{hyperref}
\usepackage{cite}
\usepackage{mathtools}
\usepackage{amsmath}
\usepackage{rotating,slashed,amsmath,charter,xcolor,catchfilebetweentags,ifluatex}

\usepackage{graphicx}
\usepackage{sidecap}

\usepackage[latin1]{inputenc} 
\usepackage{multicol}  
 
 \usepackage{titlesec}
 
 \usepackage{rotating,slashed,xcolor,amsfonts,expdlist,charter}

\numberwithin{equation}{section}

\usepackage{xcolor}
\usepackage{sectsty}

\usepackage{mdframed}
\usepackage{titletoc}

\definecolor{secnum}{RGB}{13,151,225}
\definecolor{ptcbackground}{RGB}{212,237,252}
\definecolor{ptctitle}{RGB}{0,177,235}

\titlecontents{lsection}
  [5.8em]{\sffamily}
  {\color{secnum}\contentslabel{2.3em}\normalcolor}{}
  {\titlerule*[1000pc]{.}\contentspage\\\hspace*{-5.8em}\vspace*{5pt}%
    \color{white}\rule{\dimexpr\textwidth-15.5pt\relax}{1pt}}

\usepackage{hyperref}
\hypersetup{colorlinks,bookmarksopen,bookmarksnumbered,citecolor=blus,
linkcolor=redy,pdfstartview=FitH,urlcolor=blus}
\usepackage{slashed}

\definecolor{blus}{cmyk}{1,0.9,0,0.1}
\definecolor{verdes}{cmyk}{0.99,0,0.59,0.65}
\definecolor{rossos}{cmyk}{0,1,1,0.55}
\definecolor{redy}{cmyk}{0,1,1,0.7}
\definecolor{greeny}{cmyk}{0.99,0,0.59,0.98}
\definecolor{green-go}{cmyk}{0.79,0,0.59,0.5}

\usepackage{titlesec}

\def\Lag{\mathscr{L}}

\newcommand{\beq}{\begin{equation}}
\newcommand{\eeq}{\end{equation}}

\def\hhref#1{\href{http://arxiv.org/abs/#1}{arXiv:#1}} 

 \def\Lag{\mathscr{L}}
 
\newcommand{\tmtextbf}[1]{{\bfseries{#1}}}
\newcommand{\tmtextrm}[1]{{\rmfamily{#1}}}
\def\bp{\bar M_{\rm Pl}}

\def\be{\begin{equation}}
\def\ee{\end{equation}}
\def\ba{\begin{array} }

\def\bac{\begin{array} {c}}
\def\bacc{\begin{array} {cc}}
\def\baccc{\begin{array} {ccc}}
\def\bacccc{\begin{array} {cccc}}
\def\ea{\end{array}}
\def\bea{\begin{eqnarray}}
\def\eea{\end{eqnarray}}

\definecolor{red}{rgb}{1,0,0}

\def\psl{\hbox{\hbox{${p}$}}\kern-1.9mm{\hbox{${/}$}}}
\def\dsl{\hbox{\hbox{${\partial}$}}\kern-2.2mm{\hbox{${/}$}}}
\def\Dsl{\hbox{\hbox{${D}$}}\kern-2.6mm{\hbox{${/}$}}}

\def\Lag{\mathscr{L}}

\newcommand{\gappeq}{{\rlap{{\raise}.5ex\text{\ensuremath{>}}}{{\lower}.5ex\text{\ensuremath{\sim}}}}}
\newcommand{\lappeq}{{\rlap{{\raise}.5ex\text{\ensuremath{<}}}{{\lower}.5ex\text{\ensuremath{\sim}}}}}
\newcommand{\I}{\tmtextrm{1{\kern}-.24em l}}

\begin{document}
\topmargin -1.0cm
\oddsidemargin 0.9cm
\evensidemargin -0.5cm

{\vspace{-1cm}}
\begin{center}

\vspace{-1cm}


 {\Huge \tmtextbf{ 
\color{rossos} Axion-Sterile-Neutrino Dark Matter}} {\vspace{.5cm}}\\

\vspace{1.9cm}

{\large  {\bf Alberto Salvio$^{a,b}$} and {\bf Simone Scollo$^a$ }
{\em  

\vspace{.4cm}

 	${}^a$ Physics Department, University of Rome Tor Vergata, \\ 
via della Ricerca Scientifica, I-00133 Rome, Italy\\

\vspace{0.6cm}

${}^b$ I. N. F. N. -  Rome Tor Vergata,\\
via della Ricerca Scientifica, I-00133 Rome, Italy\\ 

\vspace{0.4cm}
 
\vspace{0.2cm}

 \vspace{0.5cm}
}

\vspace{0.2cm}

}
\vspace{0.cm}

%
%
%
%
%

\end{center}

%
%
\noindent --------------------------------------------------------------------------------------------------------------------------------

\begin{center}
{\bf \large Abstract}
\end{center}

\noindent Extending the Standard Model  with three right-handed neutrinos and a simple QCD axion sector  can account for neutrino oscillations, dark matter and baryon asymmetry; at the same time, it solves the strong CP problem, stabilizes the electroweak  vacuum and can implement critical Higgs inflation (satisfying all current observational bounds). We perform here a general analysis of dark matter (DM) in such a model, which we call the $a\nu$MSM. Although critical Higgs inflation features a (quasi) inflection  point of the inflaton potential we show that DM cannot receive a contribution from primordial black holes in the $a\nu$MSM. This leads to a multicomponent axion-sterile-neutrino DM and allows us to relate the axion parameters, such as the axion decay constant, to the neutrino parameters. We include several DM production mechanisms: the axion production via misalignment and decay of topological defects as well as the sterile-neutrino production through the resonant and non-resonant mechanisms and in the recently proposed CPT-symmetric universe. 
  
  \vspace{0.4cm}

\noindent --------------------------------------------------------------------------------------------------------------------------------

\vspace{1.1cm}


\vspace{2cm}

Email: alberto.salvio@roma2.infn.it


\newpage

\tableofcontents


\section{Introduction}\label{Introduction}

 Despite the remarkable success of the Standard Model (SM),  there is no question that it  needs to be extended. The observational evidence for neutrino oscillations and DM is indeed enough to draw this conclusion.

A minimal phenomenological completion of the SM up to the Plank scale was presented in~\cite{Salvio:2015cja}, where the SM was extended to include three right-handed neutrinos with a generic flavour structure and the extra fields of the simplest invisible QCD $a$xion  model, the KSVZ one~\cite{Kim:1979if}. The model of~\cite{Salvio:2015cja}, which we refer to as the  $a\nu$MSM, not only accounts for neutrino oscillations and DM, but it can also provide the observed amount of baryon asymmetry in the universe, stabilize the electroweak (EW) vacuum, realize Higgs inflation~\cite{Bezrukov:2007ep,Bezrukov:2009db,Bezrukov:2009-2,Salvio-inf}
 and solve the strong CP problem through the Peccei-Quinn (PQ) symmetry\footnote{The strong CP problem is the fine-tuning problem of explaining why the strong interactions do not break CP, while EW ones do. Addressing this fine-tuning problem through a symmetry without doing the same with the Higgs mass and cosmological constant fine-tuning problems appears to be a logical possibility,  because the latter problems could be both addressed through anthropic arguments~\cite{Weinberg:1987dv} (unlike the strong CP one).}~\cite{Peccei:1977hh} at the same time.

In Ref.~\cite{Salvio:2018rv} it was found that Higgs inflation can be realized in its critical version~\cite{Hamada:2014iga,Bezrukov:2014bra,Hamada:2014wna} within the $a\nu$MSM: critical Higgs inflation (CHI) occurs when the SM lies extremely close to the border between the absolute stability and metastability of the EW vacuum~\cite{Buttazzo:2013uya}. CHI is particularly interesting for two reasons. One is that it can occur with a moderate, ${\cal O}(10)$, non-minimal coupling $\xi_H$ between the Higgs and the Ricci scalar. Consequently, the scale of breaking of perturbative unitarity, which was noticed in~\cite{crit}, is pushed just below the Planck scale where anyhow new physics is required to UV complete gravity. Furthermore, in Ref.~\cite{Salvio:2017oyf} it was shown that CHI, unlike standard Higgs inflation~\cite{Salvio:2015kka}, does not suffer from fine tuning in the initial conditions before inflation.  It is also interesting that one will be able test this inflationary scenario with future space-borne interferometers~\cite{Salvio:2021kya}.

So far DM in this model has been accounted for exclusively through the axion. However, the $a\nu$MSM is rich enough to contain other potential DM candidates. DM is one of the biggest mysteries in fundamental physics, it represents the majority of matter in our universe, but its nature is still unclear. Motivated by these and other  facts (see below) here we perform a general analysis of DM in the $a\nu$MSM.

We now provide an outline of this paper, which includes a  summary of the results and highlights the motivations and the original parts. 

In Sec.~\ref{model} we briefly review the $a\nu$MSM. The gauge group, ${\rm SU(3)_c\times SU(2)_{\it L}\times U(1)_{\it Y}}$, is the same as that of the SM, but the field content is extended to include the three right-handed neutrinos, a complex scalar (gauge singlets) and two Weyl fermions that are charged under the color gauge factor ${\rm SU(3)_c}$ only. The gravitational sector includes non-minimal couplings of all scalars to gravity, which allow inflation to take place. Sec.~\ref{model} also includes a discussion of the generic observational bounds that are needed for our purposes (other than the bounds related to DM, which are then discussed in the following sections).

Sec.~\ref{Axion-DM} focuses on the axion contribution to DM. As explained there, we include the contribution from both the misalignment mechanism~\cite{axionDMmis} and the decay of topological defects~\cite{axionDMstring}, which have been computed for the KSVZ model in~\cite{Ballesteros:2016xej}. The latter contribution to the DM energy density has a dependence on the quartic coupling of the extra scalar, whose value in the relevant parameter space of the  $a\nu$MSM is determined here explicitly.

Secs.~\ref{Sterile-neutrino-DM} and~\ref{A mention of the CPT symmetric case} are dedicated to  the contribution to DM due to the lightest sterile neutrino. This is a good warm dark matter candidate when its mass is around the keV. Three possible mechanisms are found. The first two are the non-resonant~\cite{Dodelson:1993je}  and resonant~\cite{Shi:1998km} production mechanisms, which occur thanks to the mixing between such sterile neutrino and the active neutrinos of the SM (see Refs.~\cite{Kusenko:2009up,Adhikari:2016bei,Boyarsky:2018tvu} for reviews). The third one takes place in a recently proposed CPT-symmetric universe~\cite{Boyle:2018tzc,Boyle:2018rgh}, where inflation and the above-mentioned mixing are not required.  For all these mechanisms we derive the contributions to the DM energy density  and the observational bounds as functions of the DM fraction $X_s$ due to the lightest sterile neutrino. Some of these functions were already known in the literature, while others are extracted here, as discussed in those sections.

Since in all the sterile-neutrino production mechanisms the masses of these neutral fermions are below the $\sim10^{14}$ GeV scale, they necessarily have a negligible impact on the running and, consequently,  the parameter space of the $a\nu$MSM  with absolute EW vacuum stability is enlarged~\cite{Salvio:2015cja}. This is because, generically, a Yukawa coupling (that is proportional to the mass of a fermion) contributes negatively to the $\beta$-function of the Higgs quartic coupling, as explained at the end of Sec.~\ref{A mention of the CPT symmetric case}. Furthermore, the presence of a sizeable sterile-neutrino contribution to DM, as we will discuss explicitly, allows to reduce the mass of the extra scalar for fixed values of its couplings and so to stabilize the EW vacuum more efficiently~\cite{RandjbarDaemi:2006gf,EliasMiro:2012ay,Salvio:2015cja,Salvio:2018rv}. All the sterile-neutrino production mechanisms, therefore, favor EW vacuum stability. This is another motivation for realising a fraction of DM through sterile neutrinos in the $a\nu$MSM.  

Yet another motivation for this work is the fact that the well-motivated presence of the axion also significantly enlarges the viable region of parameter space for sterile-neutrino DM in the $a\nu$MSM, compared to the case\footnote{This is the case e.g. in the $\nu$MSM~\cite{Asaka:2005pn,Asaka:2005an,Asaka:2006ek,Asaka:2006nq,Canetti:2012vf}, where the axion sector is absent.} $X_s=1$ (where such region is quite narrow~\cite{Abazajian:2019ejt,Perez:2016tcq}):  all observational bounds become weaker when the sterile neutrino has to account for only a fraction $X_s<1$ of DM. 

Another possible source of DM in the $a\nu$MSM could be due to primordial black holes (PBHs): CHI features a (quasi) inflection point in the inflaton potential, which has been proposed in~\cite{Garcia-Bellido:2017mdw,Ezquiaga:2017fvi,Ballesteros:2017fsr,Hertzberg:2017dkh,Motohashi:2017kbs} as a potential trigger for PBH DM production (see Ref.~\cite{Carr:2020gox} for a review). However, in Sec.~\ref{Primordial black holes as dark matter?} we show that, although this feature is qualitatively present, the $a\nu$MSM is not quantitatively able to account for any fraction of DM in the form of PBHs.

Therefore, the $a\nu$MSM leads to an axion-sterile-neutrino DM scenario, which allows us to relate the axion parameters such as the axion decay constant $f_a$ to the sterile neutrino parameters (the masses of these neutral particles and their mixing  with the active neutrinos); this provides us with an interesting link between neutrino and axion physics. The allowed parameter space for this combined axion-sterile-neutrino DM scenario is identified in Sec.~\ref{Axion-sterile-neutrino-DM} taking into account the previously discussed bounds. 

Finally, in Sec.~\ref{Conclusions} we offer our conclusions.

 \section{The $a\nu$MSM and generic observational bounds}\label{model}

We now give the details  of the $a\nu$MSM that are needed for our purposes (see Refs.~\cite{Salvio:2015cja,Salvio:2018rv} for an introduction to this model). The SM is extended with three sterile neutrinos $N_i$ and the  fields of the KSVZ axion model~\cite{Kim:1979if} (two Weyl fermions $q_1$, $q_2$ neutral under ${\rm SU(2)_{\it L}\times U(1)_{\it Y}}$ and a complex scalar $A$) .

 Correspondingly, the SM Lagrangian, $\Lag_{\rm SM}$,  is extended by adding three terms,
 \be \mathscr{L} = \Lag_{\rm SM}+\Lag_{N}+  \Lag_{\rm axion}+ \Lag_{\rm gravity},\label{full-lagrangian} \ee
 which we define in turn.  $\Lag_{N}$ represents the $N$-dependent piece:
  \be  i\overline{N}_i \dsl N_i+ \left(\frac12 N_i M_{ij}N_j +  Y_{ij} L_iH N_j + {\rm h.  c.}\right). \ee
We take the Majorana mass matrix $M$ diagonal and real, $M=\mbox{diag}(M_1, M_2, M_3),$ without loss of generality, but the Yukawa matrix $Y$ is generic.
  $\Lag_{\rm axion}$ is  the KSVZ piece:
  \be \mathscr{L}_{\rm axion} = i\sum_{j=1}^2\overline{q}_j \Dsl \, q_j +|\partial A|^2  -(y q_2A q_1 +h.c.)-\Delta V(H,A),\nonumber \ee
where $\Delta V(H,A)$ is the $A$-dependent piece of the classical potential
\be \Delta V(H,A) \equiv \lambda_A(|A|^2-  f_a^2/2)^2 + \lambda_{HA} (|H|^2-v^2)( |A|^2-f_a^2/2),\nonumber \ee
  $v\simeq 174$~GeV is the EW breaking scale and $f_a$ is the axion decay constant.
  The Yukawa coupling  $y$ is chosen real and positive without loss of generality. Finally, 
 \be \mathscr{L}_{\rm gravity} = -\left(\frac{\bp^2}{2} +\xi_H (|H|^2-v^2) + \xi_A (|A|^2-f_a^2/2)\right)R -\Lambda,
 \label{gravity-Lag} \ee
where $\bp$ is the reduced Planck mass, $R$ is the Ricci scalar,
 $\xi_H$ and $\xi_A$ are the non-minimal couplings of $H$ and $A$ to gravity and $\Lambda$ is the cosmological constant. In our model the inflaton is identified with the Higgs; it is possible to do so with $\xi_H \sim\mathcal{O}(10)$,  as discussed in Ref.~\cite{Salvio:2018rv}, when we are close to the frontier between the stability and the metastability of the EW vacuum (critical Higgs inflation).

 After EW symmetry breaking the neutrinos acquire a Dirac mass matrix
$ m_D = v Y$, which
 can  be parameterized in terms of  column vectors $m_{Di}$ ($i=1,2,3$), i.e. $m_D =\left(\begin{array}{ccc}\hspace{-0.1cm}m_{D1}\,, & \hspace{-0.2cm}m_{D2}\, ,  & \hspace{-0.2cm} m_{D3}\hspace{-0.1cm}
\end{array}\right).$
The active-neutrino masses $m_i$ ($i=1,2,3$) are obtained by diagonalizing the matrix \be m_\nu= \frac{m_{D1} m_{D1}^T}{M_1} + \frac{m_{D2} m_{D2}^T}{M_2} + \frac{m_{D3} m_{D3}^T}{M_3} . \label{see-saw}\ee
We then express $Y$ in terms of the $M_i$ and  $m_i$  as done in Refs.~\cite{Salvio:2015cja,Salvio:2018rv}.

On the other hand, the PQ symmetry breaking induced by $\langle A\rangle=f_a/\sqrt{2}$ leads to the quark  mass $M_q = y f_a/\sqrt{2}$ and the scalar squared mass
\be M_A^2 = f_a^2\left(2\lambda_A +\mathcal{O}\left(\frac{v^2}{f_a^2}\right)\right).  \label{MA1} \ee
Since $f_a\gtrsim 10^8$~GeV (see Ref.~\cite{DiLuzio:2020wdo} for a review), the $\mathcal{O}\left(v^2/f_a^2\right)$ term is very small and will be neglected.
 
 Let us now discuss the other generic observational bounds that are relevant for our purposes\footnote{See Refs.~\cite{Salvio:2015cja,Salvio:2018rv} for a  discussion of the remaining observational bounds.} (with the exception of the bounds related to DM, which will be discussed in the following sections). As far as the active-neutrinos are concerned, we have several data from oscillation and non-oscillation experiments. For example, Refs.~\cite{Esteban:2020cvm,deSalas:2020pgw}  presented some of the most recent determinations of  $\Delta m^2_{21}$ and $\Delta m^2_{3l}$ (where $\Delta m_{ij}^2 \equiv m_i^2-m_j^2$ and $ \Delta m^2_{3l} \equiv  \Delta m^2_{31} $ for normal ordering and $ \Delta m^2_{3l} \equiv  - \Delta m^2_{32} $ for inverted ordering), as well as of the active-neutrino mixing angles  and the CP phase in the Pontecorvo-Maki-Nakagawa-Sakata (PMNS) matrix. Here we take  the  currently most precise values reported in~\cite{Esteban:2020cvm,deSalas:2020pgw} for normal ordering (which is currently preferred). Regarding the SM sector, we also have to fix the values of the relevant SM couplings at the EW scale, say at the top mass $M_t\simeq 172.5$ GeV~\cite{Particle data group (top)}. We take the values computed in~\cite{Buttazzo:2013uya}, which expresses these quantities in terms of $M_t$, the Higgs mass $M_h\simeq 125.1$ GeV~\cite{Particle data group}, the strong fine-structure constant renormalized at the $Z$ mass, $\alpha_s(M_Z) \simeq 0.1184$~\cite{Bethke:2012jm} and $M_W\simeq 80.379$ GeV~\cite{Particle data group} (see the quoted literature for the uncertainties on these quantities).

 \section{Axion dark matter}\label{Axion-DM}
 
  As we will discuss, axion DM is produced in the $a\nu$MSM by two mechanisms: the misalignment one~\cite{axionDMmis} and the decay of topological defects~\cite{axionDMstring,Ballesteros:2016xej}.  In order to determine these contributions to DM the topological susceptibility $\chi$ (given in terms of   $m_a$ by $\chi=m_a^2 f_a^2$) is needed. The axion mass $m_a$ and thus $\chi$ are complicated functions of the temperature $T$. Here we use the precise calculations of $\chi$ provided by~\cite{Borsanyi:2016ksw,Petreczky:2016vrs}. 
 
 As discussed in~\cite{Ballesteros:2016xej}, the energy density $\rho_a^{\rm mis}$ due to axions produced by the misalignment mechanism contributes a fraction $\Omega_a^{\rm mis} = \rho_a^{\rm mis}/\rho_{\rm cr}$ given by\footnote{As usual $h \equiv H_0/(100 {\rm km}\,{\rm s}^{-1} {\rm Mpc}^{-1})$, where $H_0$ is the Hubble constant and $\rho_{\rm cr}$ is the critical energy density.}
 \be \Omega_a^{\rm mis}h^2= (0.12\pm0.02)\left(\frac{f_a}{1.92\times 10^{11}\mbox{GeV}}\right)^{1.165}. \label{omegaa}\ee
 Requiring that the axion energy density does not exceed the total DM energy density $\rho_{\rm DM}$ (and using $\Omega_{DM}\equiv \rho_{\rm DM}/\rho_{\rm cr} = (0.1186\pm0.0020)/h^{2}$\cite{Tanabashi:2018oca}) we find the upper bound
	\begin{equation}
		f_{a}\lesssim2\times 10^{11}\;\rm{GeV}\qquad  (\mbox{from misalignment}).\label{bound_axion}
	\end{equation}
 Higgs inflation features a high reheating temperature,  $T_{\rm RH}\gtrsim 10^{13}$~GeV, thanks to the sizable couplings between the Higgs and other SM particles~\cite{Bezrukov:2008ut,BellidoReh}.  Thus $T_{\rm RH} \gg f_a$ and the PQ symmetry is restored after inflation in the $a\nu$MSM.
 
Therefore, here axion DM is also produced through decays of topological defects, which leads to a contribution $\rho^{\rm string}_{a}\equiv \Omega^{\rm string}_{a} \rho_{\rm cr}$ to the energy density, which in our model is  given by~\cite{Ballesteros:2016xej}
\be \Omega^{\rm string}_{a}h^2= 0.37^{+0.3}_{-0.2}\left(\frac{f_a}{1.92\times 10^{11}\mbox{GeV}}\right)^{1.165} \frac{\ln\left(f_at_{co}\sqrt{\lambda_A/\zeta}\right)}{50}. \label{omegaas}\ee
In Eq.~(\ref{omegaas}) the time $t_{co}$ can be determined in terms of $m_a$ by 
\be \frac{2\pi\epsilon_a}{t_{co}} = m_a(t_{co}) \label{tco} \ee and the numerical simulations of~\cite{Kawasaki:2014sqa} give $\epsilon_a=4\pm 0.7$ and $\zeta=1\pm 0.5$. 

Therefore, the time $t_{co}$ can be  computed by using the precise calculations of $\chi$. However, since $\Omega^{\rm string}_{a}$ depends on $t_{co}$ only logarithmically we can, as we explain now, simply estimate $t_{co}$ by using the dilute instanton gas approximation, which gives  a power-law temperature dependence of $\chi$,
\be \chi(T)=\chi_{0}\left(\frac{T_{\rm QCD}}{T}\right)^{n}, \label{DIGA}\ee 
where $T_{\rm QCD}$ is the temperature of the QCD confining phase transition ($T_{\rm QCD}\simeq 157$~MeV), $n\simeq8.16$ and $\chi_{0}\simeq0.0216\;{\rm fm^{-4}}\simeq(75.6\;{\rm MeV})^4$~\cite{Borsanyi:2016ksw}.
Here our treatment starts to diverge from that of~\cite{Ballesteros:2016xej} because the quartic coupling $\lambda_A$ does not need to be tiny in our model (unlike in~\cite{Ballesteros:2016xej}). First, note that in the radiation dominated era the Friedmann equation can be written in the form 
\be t^{-2}=\frac{2\pi^{2}}{45\bp^2}g_{*}(T)T^{4}, \label{Feq}\ee
 where $g_{*}$ is the effective  number of relativistic species. Using this result and Eqs.~(\ref{tco}) and~(\ref{DIGA}) one finds 
\begin{equation}
	T_{co}=\bigg(\frac{45\bp^{2}\chi_{0}T^{n}_{\rm QCD}}{8\pi^{4}\epsilon^{2}_{a}f^{2}_{a}g_{*}}\bigg)^{\frac{1}{4+n}}\simeq  \;{\rm GeV}\bigg(\frac{2\times10^{11}\;{\rm GeV}}{f_{a}}\bigg)^{\frac{2}{4+n}}g_{*}^{-\frac{1}{4+n}},
\end{equation}
so
\begin{equation}
	T_{co}\simeq 0.8\;{\rm GeV}\bigg(\frac{2\times10^{11}\;{\rm GeV}}{f_{a}}\bigg)^{\frac{2}{4+n}}, \label{TcoComp}
\end{equation}
where we used well-known determinations of $g_*$ in the SM (see e.g.~\cite{Borsanyi:2016ksw}) and the fact that the contributions of the extra particles beyond the SM to $g_*$ are negligible at those temperatures. As a check of this result  note that the power-law temperature dependence of $\chi$ fits reasonably well the full lattice results already from temperatures of order of few hundreds of MeV (see Fig.~2 of~\cite{Borsanyi:2016ksw}). Now, using again the Friedmann equation in~(\ref{Feq}) we have
 \begin{equation}
	t_{co}\simeq4\times 10^{-7}\;{\rm s}\;\bigg(\frac{f_{a}}{2\times10^{11}\;{\rm GeV}}\bigg)^{\frac{4}{4+n}}. \label{tcoComp}
\end{equation}
We can equivalently use Eq.~(\ref{TcoComp}) or Eq.~(\ref{tcoComp}) to estimate the argument of the logarithm in~(\ref{omegaas}) because~(\ref{tco}) and~(\ref{DIGA}) tell us
\begin{equation}
	f_{a}t_{co}\sqrt{\lambda_A/\zeta}=\frac{2\pi\epsilon_{a} f^{2}_{a}\sqrt{\lambda_A}}{\sqrt{\chi_{0}\zeta}}\bigg(\frac{T_{co}}{T_{\rm QCD}}\bigg)^{\frac{n}{2}}.
\end{equation}

\begin{figure}[t]
\begin{center}
 \includegraphics[scale=1.1]{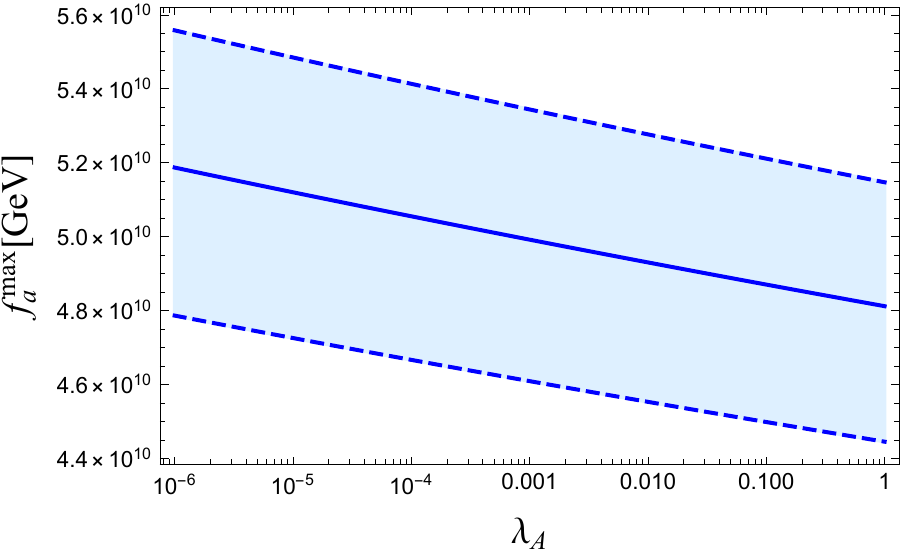}

 \end{center} 
   \caption{\em Dependence on $\lambda_A$ of the maximal value of $f_a$ obtained by requiring that DM is not overproduced in the $a\nu$MSM. The width of the band corresponds to the uncertainties in Eqs.~(\ref{omegaa}) and~(\ref{omegaas}).}
\label{famax}
\end{figure}

Therefore, Eq.~(\ref{TcoComp}) or Eq.~(\ref{tcoComp}) allows us to estimate $\Omega^{\rm string}_a$ for each value of $f_a$ and $\lambda_A$.  So fixing $\lambda_A$ we obtain a maximal value of $f_a$, which we call here $f^{\rm max}_a$, from the requirement that the axion energy density does not exceed the total DM energy density $\rho_{\rm DM}$, namely 
\be  \Omega_a^{\rm mis}+ \Omega^{\rm string}_{a} \leq \Omega_{\rm DM}, \ee 
where $\Omega_{\rm DM} \equiv \rho_{\rm DM}/\rho_{\rm cr}$.
If one considers for example $\lambda_A =0.1$  the argument of the logarithm in~(\ref{omegaas})  becomes
\begin{equation}
	f_{a}t_{co}\sqrt{\lambda_A/\zeta}\simeq 4\times10^{28}\bigg(\frac{f_{a}}{2\times10^{11}\;{\rm GeV}}\bigg)^{\frac{8+n}{4+n}}.
\end{equation}
and so 
\begin{equation}
	f_{a}\leq f^{\rm max}_a\simeq 5\times10^{10}\;{\rm GeV} \qquad (\mbox{for}\,\,  \lambda_A = 0.1),
\end{equation}
which is significantly lower than the pure misalignment bound in~(\ref{bound_axion}). In Fig.~\ref{famax} we show how $f^{\rm max}_a$ depends on $\lambda_A$.

 \section{Sterile-neutrino dark matter}\label{Sterile-neutrino-DM}

 Another important source of DM in the $a\nu$MSM is  the sterile neutrino $\tilde N_1$ with the smallest mass $m_s$. Generically, this is not exactly $N_1$ due to an active-sterile neutrino mixing. Indeed,  in an inflationary\footnote{In Sec.~\ref{A mention of the CPT symmetric case} we will discuss the recently proposed CPT-symmetric universe of~\cite{Boyle:2018tzc,Boyle:2018rgh} where inflation is not required.} universe the production of this particle occurs through its mixing with the active neutrinos of the SM. Such mixing is described by three (generically complex) quantities $\theta_{\alpha1}$, where $\alpha$ represents the flavour of the active neutrino $\nu_\alpha$ ($\alpha=e,\mu,\tau$). The $\theta_{\alpha1}$ are the elements of the matrix $\Theta \equiv m_D M^{-1}$.
 It is convenient to introduce a total mixing parameter $\theta$ defined by~\cite{Boyarsky:2018tvu}
 \be \theta^2\equiv \sum_{\alpha=e,\mu,\tau}|\theta_{\alpha1}|^2. \ee 
 
 The mixing with the active neutrinos leads to the production of sterile neutrinos in the early universe in two ways. One is provided by the oscillations between active and sterile states. Furthermore, sterile neutrinos are produced in scatterings. Though this production mechanism is ``thermal" in the sense that the $\tilde N_1$ are produced in scatterings in a thermal plasma, generically these particles are not in thermal equilibrium because of their tiny couplings. 
As we will discuss in the following subsections, the energy density $\rho_s$ of the lightest sterile neutrino  can generically account for a non-negligible fraction $X_s$ of the total DM abundance, i.e. $\Omega_s = X_s \Omega_{\rm DM}$, where $\Omega_s = \rho_s/\rho_{\rm cr}$ and $0\leq X_s\leq 1$.  
 
 In order to identify the allowed regions of the parameter space it is necessary to have the observational bounds for an arbitrary value of $X_s$. Of course, the bounds will be generically weaker for $X_s<1$ than for $X_s=1$, but we want to know how they change varying $X_s$.

  First, a fermionic DM candidate is subject to a phase-space lower bound on its mass (a.k.a the Tremaine-Gunn bound~\cite{Tremaine:1979we}) that is related to Pauli's exclusion principle\footnote{See~\cite{Gorbunov:2008ka} for a study of this bound when sterile neutrinos account for the whole DM.}. The strongest information comes from the dwarf spheroidal galaxies (dSphs), which are the most compact DM-dominated objects observed so far.
   In objects of this sort the dynamics of the DM particles can be characterized  by some coarse-grained primordial phase-space density ${\cal D}$ and the one-dimensional velocity $\sigma$: see~\cite{Boyanovsky:2007ay} for a detailed discussion. Since the coarse-grained  phase-space density either remains constant or diminishes, today we have~\cite{Boyanovsky:2007ay} (see also~\cite{deVega:2009ku})
  \be \frac{\rho_s}{\sigma^3}\leq 3^{3/2} m_s^4 {\cal D}. \ee
Therefore, writing $\rho_s = X_s \rho_{\rm DM}$ we obtain the lower mass bound
\be m_s\geq \left(\frac{X_s \rho_{\rm DM}}{3^{3/2} {\cal D}\sigma^3}\right)^{1/4}.\ee
This result tells us that the phase-space bound is rescaled towards smaller values  by $X_s^{1/4}\leq 1$. A recent publication~\cite{Savchenko:2019qnn}, which we use here, has set the phase-space bound $m_s \gtrsim 190$~eV at $2\sigma$ level for $X_s=1$. It is worth mentioning that an axion-like particle with mass around the $10^{-22}$~eV scale (which is not the QCD axion of the  $a\nu$MSM) can also be constrained with phase space data~\cite{deVega:2014wya}. 

Other important bounds on sterile neutrino DM come from the search of X-rays produced by the radiative decay $\tilde N_1\to \gamma \nu_\alpha$~\cite{Pal:1981rm,Barger:1995ty}. Since the differential flux produced by the decay of sterile neutrinos depends on the
product $X_s \sin^2(2\theta)$, for each chosen value of $m_s$, the upper limit on $\sin^2(2\theta)$ must weaken by decreasing $X_s$, rescaling exactly as $1/X_s$ (see also Ref.~\cite{Benso:2019jog} for a related study). The most recent publications providing this type of X-ray bounds used the data collected by the NuSTAR satellite, a space-based X-ray telescope, observing the Milky Way (see~\cite{Boyarsky:2018tvu} for a review)  and, more recently, the Andromeda galaxy (M31)~\cite{Ng:2019gch}. We will take these bounds into account  in Sec.~\ref{Axion-sterile-neutrino-DM}.

Now we describe in turn various production mechanisms for $\Omega_s$ as a function of $X_s$.

 \subsection{Non-resonant production}\label{Non-resonant production}
  
  One important production mechanism of sterile neutrino is the  Dodelson-Widrow (DW) mechanism~\cite{Dodelson:1993je}, which we will refer to as the non-resonant production. Precise calculations of $\Omega_s$ within this mechanism lead to~\cite{Abazajian:2005gj,Abazajian:2017tcc} 
\begin{equation}
	m_{s}\simeq3.28 \times\rm keV \bigg(\frac{\sin^2(2\theta)}{10^{-8}}\bigg)^{-0.615}\bigg(\frac{\Omega_s}{0.26}\bigg)^{0.5}\bigg\lbrace0.547\times erfc\bigg[-0.969\bigg(\frac{\it T_{\rm QCD}}{157\;\rm MeV}\bigg)^{2.15}\bigg]\bigg\rbrace, \label{DWrel}
\end{equation}
where erfc is the complementary error function. We have used here a normalization such that the argument of the curly bracket in~(\ref{DWrel}) equals 1 for $T_{\rm QCD}\simeq157$~MeV.

\vspace{0.4cm}

In addition to the phase-space and X-ray bounds already discussed, sterile-neutrino DM is also subject to  structure-formation bounds. This is because the typical sterile-neutrino momentum distribution exhibit a free-streaming length in the early universe, which modifies the formation of structures. This type of bounds are affected by considerable uncertainties related to, among other things, the simulation of non-linear structure formation as well as the difficulty to observe small scale structures. Furthermore, this structure-formation bound depends on the specific sterile-neutrino production mechanism one considers.
For the non-resonant production a study for generic $X_s$ has, however, already been performed in~\cite{Palazzo:2007gz}. 
To have an idea of the orders of magnitude, Ref.~\cite{Palazzo:2007gz} found a bound on $m_s$ that is about 10~keV for $X_s=1$ and 1~keV for $X_s=0.1$ and so typically stronger than the phase-space bound.

\subsection{Resonant production}\label{Resonant production}

The second mechanism to produce sterile-neutrino DM is a resonantly enhanced version of the DW mechanism, which relies on a non-vanishing lepton asymmetry $L$~\cite{Shi:1998km,Fuller2} and is based on the Mikheyev-Smirnov-Wolfenstein effect~\cite{Wolfenstein:1977ue} (see Ref.~\cite{Ghiglieri:2015jua,Venumadhav:2015pla,Bodeker:2020hbo} for more recent and precise calculations). In practice the effective mixing in the plasma gets enhanced by $L$, such that the abundance of active neutrinos allows to create sterile neutrinos more efficiently. This asymmetry can be generated dynamically by the heavier sterile neutrinos $N_2$ and $N_3$ if their masses are around the GeV scale~\cite{Laine:2008pg,Canetti:2012kh,Eijima:2020shs}. Moreover, $N_2$ and $N_3$ provide a mechanism to generate baryon asymmetry through a different version of leptogenesis~\cite{Akhmedov:1998qx,Drewes:2012ma}.

The literature so far focused on the case in which all DM is due to $\tilde N_1$ (i.e. $X_s =1$), but in our model the axion also contributes to DM so we need to find more general formul\ae~that hold  for arbitrary $X_s$. 
In the non-resonant DW mechanism the quantities $\Omega_s$, $\theta$ and $m_s$ are related by Eq.~(\ref{DWrel}), which has the form $f(\Omega_s, \theta, m_s) = 0$, such that for each fixed value of $\Omega_s$ the DW mechanism is represented by a line in the $(\theta, m_s)$ plane. In the resonant production this function acquires an extra dependence on $L$, i.e.  $f(\Omega_s, \theta, m_s, L) = 0$, and the allowed region in the $(\theta, m_s)$ plane is promoted to a band, which is limited by the DW line $f(\Omega_s, \theta, m_s, 0) = 0$. There exists another bound on this band, $f(\Omega_s, \theta, m_s, L_{\rm max}) = 0$, where $L_{\rm max}$ is the maximal value of $L$ allowed by observations: the values $L>L_{\rm max}$   are ruled out because they would excessively change   the abundances of light elements produced during Big Bang Nucleosynthesis (BBN)~\cite{Serpico:2005bc}. To obtain this bound explicitly for each value of $X_s$ let us observe that the (dimensionless) yield  $Y_s\equiv n_s/s$, where $n_s$ is the  sterile neutrino density and $s$ is the entropy density, is related to $\Omega_s$ through
\be \Omega_s =\frac{m_s n_s}{\rho_{\rm cr}} = \frac{m_s Y_s}{\rho_{\rm cr}/s} \ee
so
\be m_s =\frac{X_s\Omega_{\rm DM} \rho_{\rm cr}/s}{Y_s}. \ee
Note that if we approximate $Y_s$  as a function of\footnote{In this case one neglects the dependence on $m_s/T$, where $T$ is the photon temperature. This is justified as the resonant production of sterile neutrinos occurs at $T\sim 200$~MeV and $m_s\sim$~keV~\cite{Eijima:2020shs}, so $m_s/T\sim 10^{-5}$.} 
$\theta$ and $L$ only we reproduce the linear dependence of $\Omega_s$ on $m_s$ found in~\cite{Shi:1998km}.
Using known results of the literature (see Ref.~\cite{Boyarsky:2018tvu} for a review) one obtains, within this approximation
\be m_s\gtrsim  \frac{X_s\Omega_{\rm DM} \rho_{\rm cr}/s}{Y_s(\theta,L_{\rm max})}.\ee
This formula tells us that the above-mentioned BBN bound in the resonant production band in the $(\theta, m_s)$ plane is rescaled towards smaller values of $m_s$ by $X_s\leq 1$.
 
\section{Sterile-neutrino dark matter in a CPT-symmetric universe}\label{A mention of the CPT symmetric case}

It was recently pointed out that another mechanism to produce sterile neutrino DM is present if one constructs a CPT-symmetric universe~\cite{Boyle:2018tzc,Boyle:2018rgh} in the absence of inflation: the universe before the Big Bang is the CPT reflection of the universe after the Big Bang, so that the time evolution of the universe does not spontaneously violate CPT.  In this scenario a sterile-neutrino cosmic abundance is produced according to late-time comoving observers like us just because the vacuum is time dependent. Therefore, unlike the production mechanisms of Sec.~\ref{Sterile-neutrino-DM}, a mixing of the sterile neutrino $\tilde N_1$ responsible for DM and the active neutrinos is not necessary. One can, therefore, set this mixing to zero requiring the theory to be invariant under a $Z_2$ symmetry acting on $\tilde N_1$, which also makes $\tilde N_1$ exactly stable. As a result, the sterile neutrinos produced through this mechanism can easily avoid the X-ray bounds discussed in Sec.~\ref{Sterile-neutrino-DM}.

In our model inflation can occur and can be triggered by the Higgs, therefore, we do not perform a general study of this possibility\footnote{See Ref.~\cite{Duran:2021wao} for a recent generalization of the results in~\cite{Boyle:2018tzc,Boyle:2018rgh} to non-standard, but also CPT-symmetric early universe cosmologies.}. However, it is interesting to see how the calculations of~\cite{Boyle:2018tzc,Boyle:2018rgh} change in the presence of another DM component, which in our case is due to the axion.

As shown in~\cite{Boyle:2018rgh}, assuming that this production mechanism occurs in the radiation dominated era, the yield of the sterile-neutrino can be expressed in terms of its mass $m_s$:
\be Y_s = \frac{3I}{2\pi^2}\left(\frac{15}{g_*}\right)^{1/4}\left(\frac{m_s}{\hat \mu}\right)^{3/2}, \label{YsCPT}\ee 
where 
\be I \equiv \frac1{2\pi^2}\int_0^\infty dx x^2\left[1-\sqrt{1-e^{-x^2}}\right] \simeq 0.01276 \ee
and $\hat \mu \simeq 5.966 \times 10^{18}$~GeV. In this case the predicted sterile-neutrino contribution to the DM energy density is 
\be X_s \rho_{\rm DM} = m_s n_s=m_s Y_s s.	\label{rhos}\ee	
Using the known value of $\rho_{\rm DM}/s$ we find the sterile-neutrino mass that is required to account for a fraction $X_s$ of the DM abundance:
\be m_s \simeq 4.8 \times 10^8~\mbox{GeV}~X_s^{2/5} \left(\frac{g_*}{g_*^{\rm SM}}\right)^{1/10}, \label{msCPT} \ee
where  $g_*^{\rm SM}$ is the effective number of relativistic degrees of freedom in the SM ($g_*^{\rm SM}\simeq 106.75$ for $T\gg 100$~GeV). We note a  dependence on $X_s$ and a (milder) dependence on $g_*$. 
We also observe that the phase-space lower bound discussed in Sec.~\ref{Sterile-neutrino-DM} is always satisfied down to negligibly small values of $X_s$.
 
 \vspace{0.4cm}
 
 Note that in all the sterile-neutrino production mechanisms that we have discussed in this section and Sec.~\ref{Sterile-neutrino-DM} $m_s$ is generically well below (at least six orders of magnitude) the $10^{14}$~GeV scale. Then from Eq.~(\ref{see-saw}), using the observational bounds on $m_i$ and $v$, it follows that the impact on the RGEs (see Appendix~\ref{RGEs}) of the Yukawa couplings $Y_{ij}$ is generically negligible compared to the other contributions in Appendix~\ref{RGEs}.  This is a good thing because the $Y_{ij}$ contribute negatively to the $\beta$-function  $\beta_{\lambda_H}^{(1)}$ of the Higgs quartic coupling. Moreover, when $\tilde N_1$ gives a sizeable contribution to DM the value of $f_a$  required to reproduce the observed DM abundance decreases, as clear from Sec.~\ref{Axion-DM}, then so does $M_A$ (cf.~Eq.~(\ref{MA1})). Therefore, the extra scalar starts stabilizing the EW vacuum from smaller energies~\cite{RandjbarDaemi:2006gf,EliasMiro:2012ay,Salvio:2015cja,Salvio:2018rv}. It follows that requiring the sterile neutrino to contribute to DM naturally favors EW vacuum stability.
 
\section{Primordial black holes as dark matter?}\label{Primordial black holes as dark matter?}

PBHs may be generated if the curvature power spectrum $P_\mathcal{R}$ has a peak of order $\sim 10^{-2}$~\cite{Hertzberg:2017dkh}, about seven orders of magnitude larger than at $\sim60$ e-folds before the end of inflation. An enhancement of $P_\mathcal{R}$  generically occurs when the inflaton potential features a (quasi) inflection point\footnote{See also Refs.~\cite{Kohri:2007qn,Kohri:2012yw} for earlier works on PBH production in inflationary models.}~\cite{Garcia-Bellido:2017mdw,Ezquiaga:2017fvi,Ballesteros:2017fsr,Hertzberg:2017dkh,Motohashi:2017kbs}. This is the case in CHI~\cite{Hamada:2014iga,Bezrukov:2014bra,Hamada:2014wna,Garcia-Bellido:2017mdw,Salvio:2017oyf,Ezquiaga:2017fvi}, but in order to see if $P_\mathcal{R}$ reaches the required order of magnitude in the $a\nu$MSM a study of this quantity together with other observables is required. We  perform such study in this section.

As discussed e.g. in~\cite{Salvio:2017oyf}, studying Higgs inflation in the unitary gauge, the potential of the canonically normalized Higgs field $\phi'$ is given by
\begin{equation} U_H\equiv \frac{V_H}{\Omega_H^4}=\frac{\lambda_H\phi(\phi')^4}{4(1+\xi_H\phi(\phi')^2/\bp^2)^2},\label{UH}  \end{equation}
where $V_H=\lambda_H \phi^4/4$, $\phi$ is the Higgs field non-minimally coupled to gravity, which is related to $\phi'$ through 
\begin{equation} \frac{d\phi'}{d\phi}= \Omega_H^{-2} \sqrt{\Omega_H^2 +\frac{3\bp^2}{2}  \left(\frac{d\Omega_H^2}{d\phi}\right)^2 } ,\label{phi'}\end{equation}
and $\Omega_H^2$ is defined by 
\begin{equation}  \Omega_H^2\equiv 1+\frac{2\xi_H |H|^2}{\bp^2}. \label{transformation}\end{equation}

In a spatially flat Friedmann-Robertson-Walker geometry the equations for the spatially homogeneous field $\phi'(t)$ and the cosmological scale factor $a(t)$ are  
 \be \ddot\phi' +\frac{\sqrt{3\dot\phi'^2+6U_H}}{\sqrt2\bp}\dot\phi' +\frac{dU_H}{d\phi'}  = 0\label{eq-k=0}\ee
 and 
\be H_I^2=\frac{ \dot\phi'^2+2U_H}{6 \bar M_{\rm Pl}^2},  \label{EE1}\ee
where a dot represents the derivative with respect to cosmic time $t$ and $H_I\equiv \dot a/a$. Inflation in general takes place when\footnote{As usual, the expansion of the universe is nearly exponential for $\epsilon<1$ and becomes exactly exponential as $\epsilon\to 0$.}
\be \epsilon \equiv -\frac{\dot H_I}{H_I^2}  < 1. \ee 
 Moreover, when
\be \delta \equiv  - \frac{\ddot\phi'}{H_I \dot\phi'} \ee
is small one can neglect the inertial term in the inflaton equation (\ref{eq-k=0}) and reduce the problem to a single first order differential equation, leading to the useful slow-roll approximation where
the parameters
\be \epsilon_H \equiv\frac{\bp^2}{2} \left(\frac{1}{U_H}\frac{dU_H}{d\phi'}\right)^2, \quad \eta_H \equiv \frac{\bp^2}{U_H} \frac{d^2U_H}{d\phi'^2} \label{epsilon-def}\ee
are  small. These slow-roll functions can be constructed through the more general ``horizon flow functions" of Ref.~\cite{Schwarz:2001vv}.

\begin{figure}[t]
\begin{center}
 \includegraphics[scale=0.5]{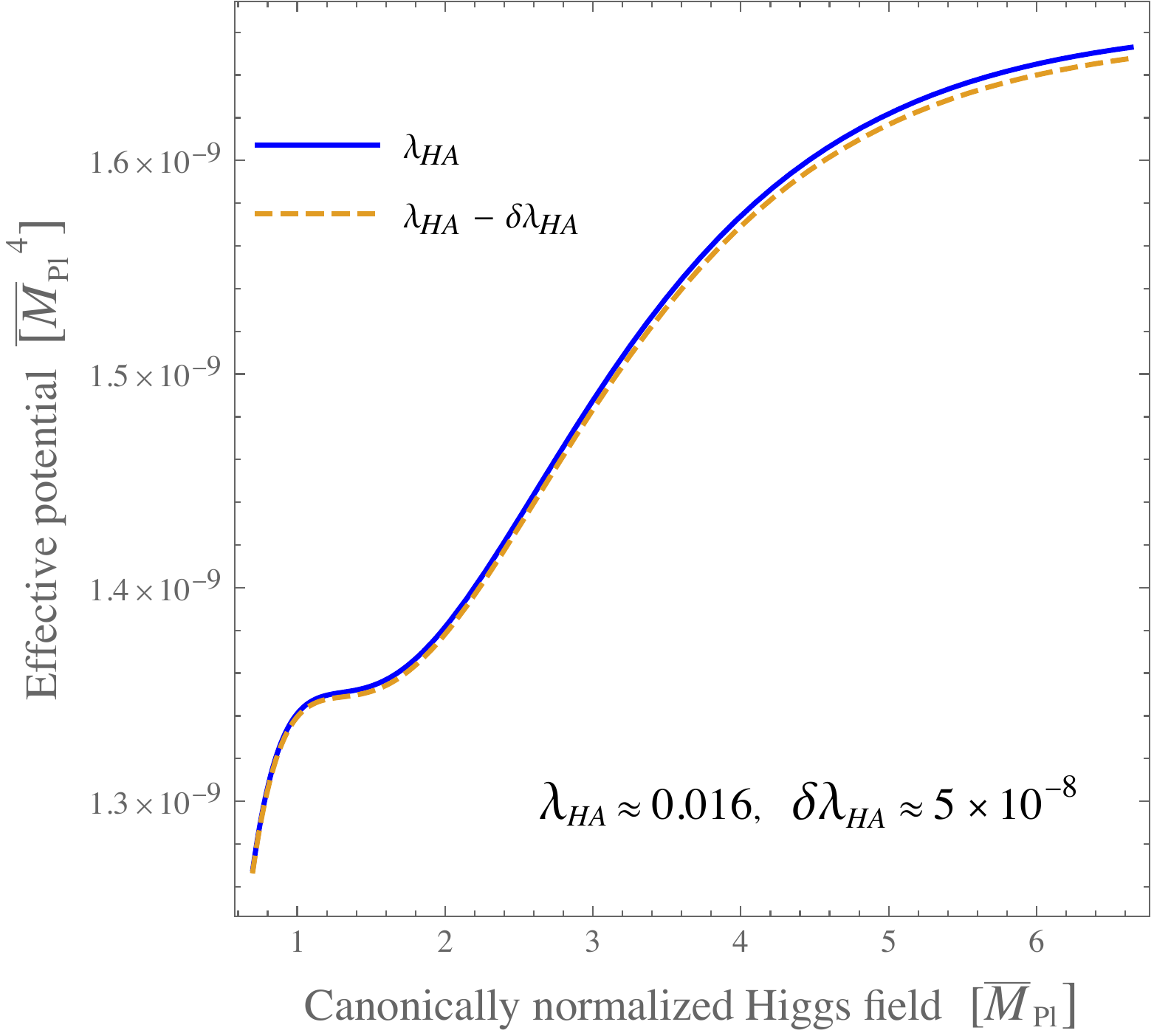}

 \end{center}
   \caption{\em The typical shape of the effective potential as a function of the canonically normalized Higgs field  close to criticality (in this plot we approach the critical regime by varying $\lambda_{HA}$). }
\label{UH}
\end{figure}

The number of e-folds is defined by
\be N_e  \equiv \int_{t_b}^{t_e} dt \,  H_I(t), \ee 
where $t_e$ is the time at the end of inflation and $t_b$ is the time when the various inflationary observables such as  $P_R$, the corresponding spectral index $n_s$ and the tensor-to-scalar ratio $r$ are determined through observations. In the slow-roll approximation $N_e$ is expressed as a function of the field $\phi_b'$ (at $t_b$) rather than as a function of time,

\begin{equation}N_e=\int_{\phi'_e}^{\phi'_{\rm b}}\frac{U_H}{\bp^2}\left(\frac{dU_H}{d\phi'}\right)^{-1}
d\phi',
\label{e-folds}\end{equation}
where $\phi'_e$ is the field value at the end of inflation, and $P_{R}$ at $\phi_b'$ can be computed through \begin{equation}P_{R}= \frac{U_H/ \epsilon_H}{24\pi^2 \bp^4}.\label{PRsr} \end{equation}

At quantum level these inflationary formul\ae~remain approximately valid except that one must consider $\lambda_H$ and $\xi_H$ as functions of $\phi'$. In defining these functions there are well-known ambiguities~\cite{Bezrukov:2009db,Bezrukov:2009-2, Bezrukov:2014bra,Bezrukov:2014ipa,Bezrukov:2017dyv}.
Here we adopt the quantization used in~\cite{Salvio:2018rv}, which can be embedded in a UV completion of gravity~\cite{Salvio:2014soa,Salvio:2016vxi,Salvio:2017qkx,Salvio:2019ewf,Salvio:2019wcp} (see Refs.~\cite{Salvio:2018crh,Salvio:2020axm} for reviews). In this approach the $\phi'$-dependence of $\lambda_H$ and $\xi_H$ is obtained by solving the RGEs given in Appendix~\ref{RGEs} as explained in~\cite{Salvio:2018rv}.
The typical shape of the effective inflationary potential (close to criticality) computed in this way is the one shown in\footnote{In that figure we chose as an example the input values  $M_1 = 10^{11}\,{\rm GeV}$, $M_2 = 6.4	 \times 10^{13}\,{\rm GeV}$, $M_3 > \bp$, 
    $f_a \simeq 2.5 \times 10^{10}\,{\rm GeV}$, $\lambda_A(M_A) \simeq 0.1$, $y(M_A)\simeq  0.1$, $\xi_H(M_A)\simeq 14$ and $\xi_A(M_A)\simeq -2.6$.} Fig.~\ref{UH}

\begin{figure}[t]
\begin{center}
 \includegraphics[scale=0.44]{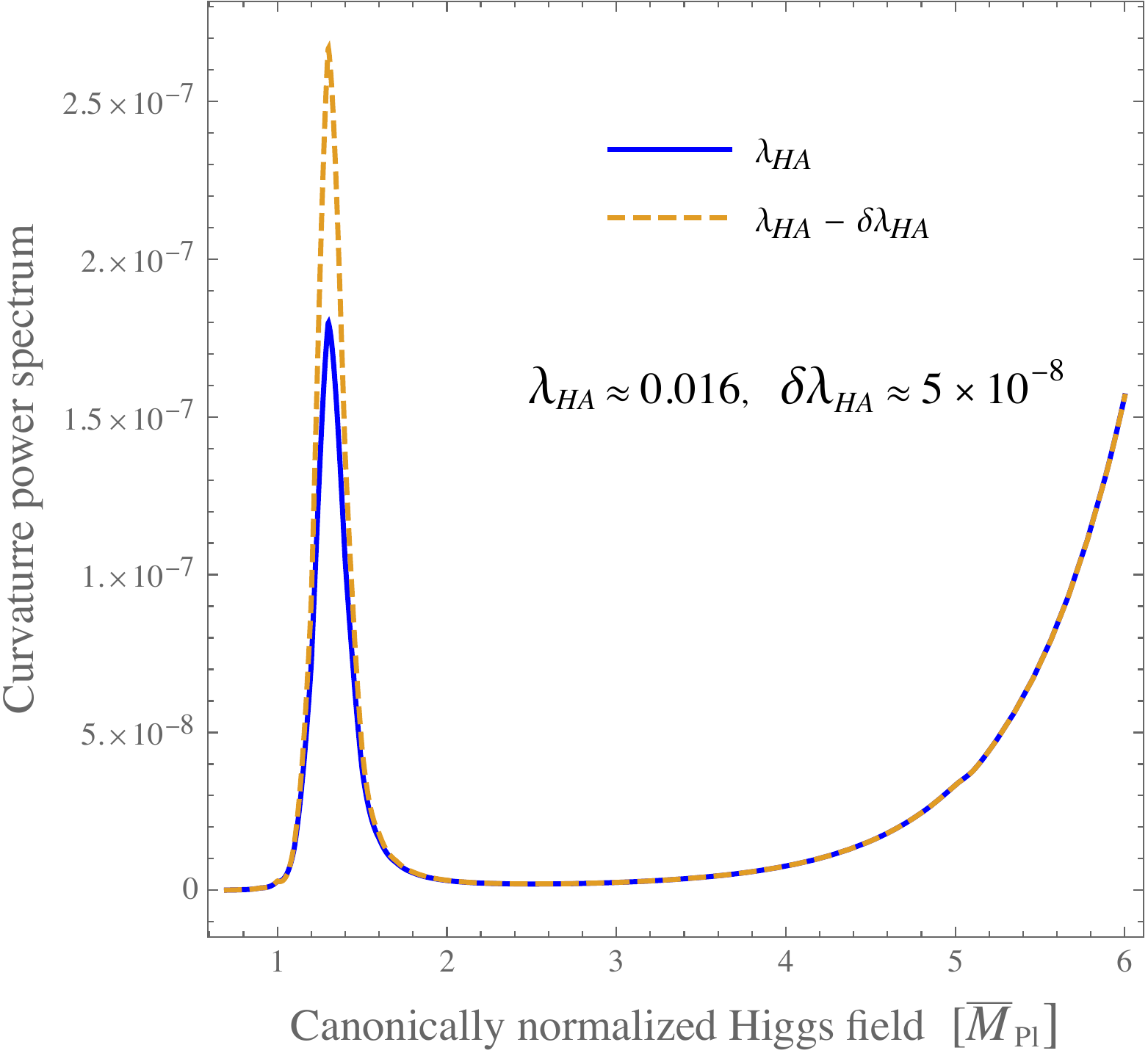}  \hspace{2.5cm} 
  \includegraphics[scale=0.39]{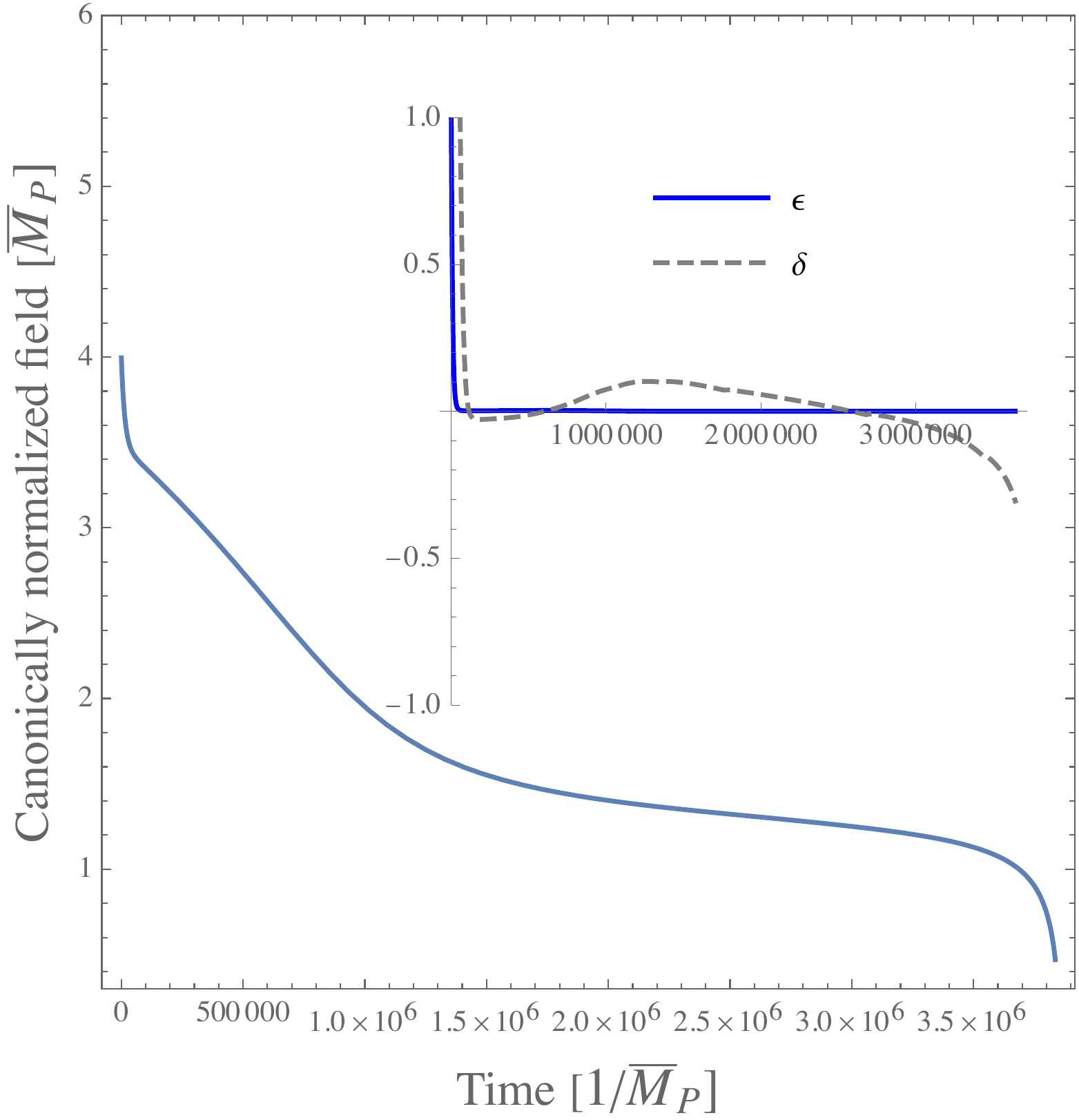}
 	 
 \end{center}
   \caption{\em The curvature power spectrum and the canonically normalized (Higgs) field  close to criticality (which we approach by varying $\lambda_{HA}$). The corresponding values of $\epsilon$ and $\delta$ are shown as well in an inset of the right plot. The parameters  are set as in Fig.~\ref{UH}. }
\label{PBHs}
\end{figure}

We find that the above-mentioned requirement to generate PBHs is never satisfied  so PBHs cannot contribute to DM in our model. The reason is the following. Although $P_\mathcal{R}$  does have a peak at a time after inflation as a consequence of the inflection point, its height  is several orders of magnitude smaller than $10^{-2}$ when one requires a plausible number of e-folds. This situation is illustrated in Fig.~\ref{PBHs}. In that figure we approach criticality by varying $\lambda_{HA}$, but varying other parameters leads to  similar situations. We find that the slow-roll approximation is still reasonably good to give at least the order of magnitude of $P_\mathcal{R}$  because both $\epsilon$ and $\delta$   are well-below 1 around the peak of the power spectrum as shown in that figure. Indeed, the number of e-folds for $\lambda_{HA}$ and $\lambda_{HA}-\delta\lambda_{HA}$ are $N_e\simeq 65$ and $N_e\simeq 71$, respectively, while the corresponding values computed with the slow-roll approximation is reasonably close ($N_e\simeq 63$ and $N_e\simeq 70$, respectively). Although the height of the peak of $P_\mathcal{R}$ does increase by approaching criticality, it does so at the price of increasing $N_e$ above the bound of~\cite{Liddle:2003as}: already for a pretty low peak of order $10^{-7}$ the number of e-folds is starting to be significantly above $\sim 60$. The more we approach criticality the larger $N_e$ becomes.

In Fig.~\ref{PBHs} we set the parameters in a way to reproduce the observed neutrino oscillations, have a stable  EW vacuum, a viable inflation and baryogenesis through leptogenesis. When all these requirements are satisfied we always find that PBHs cannot contribute to DM in our model. Essentially the reason is that the shape of the potential, although apparently  able to produce PBH DM, it does not have the right quantitative features to do so.

 \section{Axion-sterile-neutrino dark matter}\label{Axion-sterile-neutrino-DM}
 
 Having established that the only source of DM in the $a\nu$MSM are axions and sterile neutrinos, we now identify the allowed parameter space in a combined axion-sterile-neutrino DM scenario, taking into account all the previously discussed bounds. 
 
 Note that in our model $X_s$ can then be expressed as 
 \be X_s=1- \frac{\Omega_a^{\rm mis}+\Omega_a^{\rm string}}{\Omega_{\rm DM}},\label{Xsa}\ee
 which relates $X_s$ and $f_a$. We recall that $\Omega_a^{\rm string}$ also depends on $\lambda_A$ (see Eq.~(\ref{omegaas})). 
 
\begin{figure}[t]
    \centering
	\includegraphics[scale=.9]{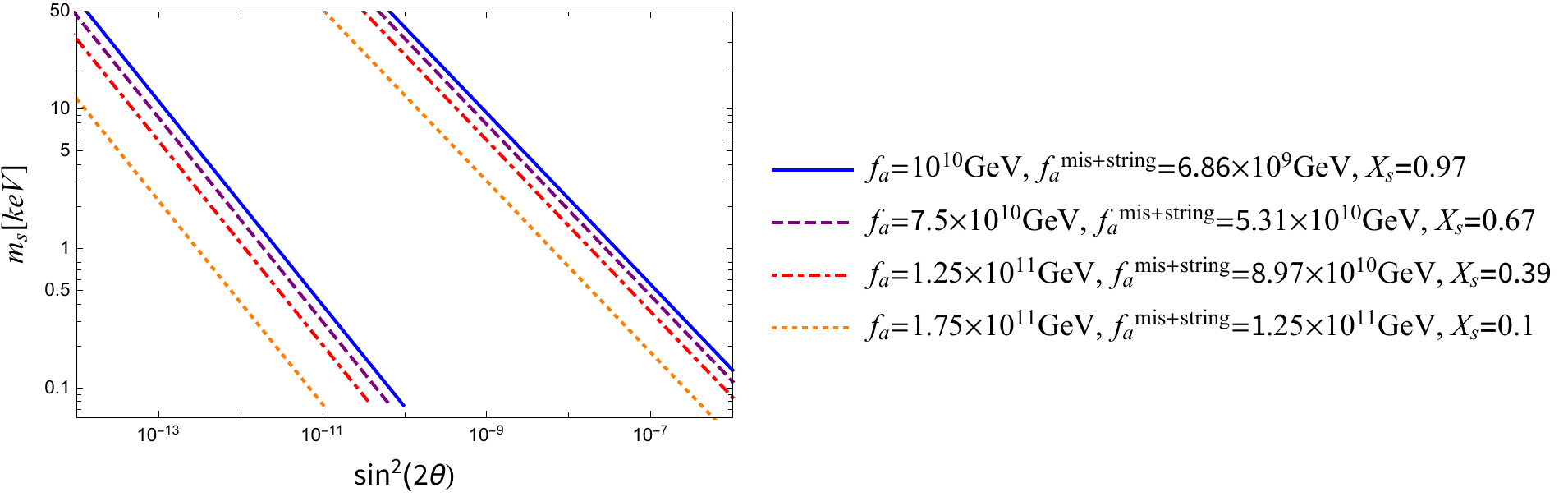}
	\caption{\em Sterile-neutrino production range as the axion decay constant changes. 
		For the first value of the axion decay constant, $f_a$, we only take into account the misalignment mechanism for axion production.  For the second value, $f_{a}^{\rm mis+string}$, we take into account both the misalignment mechanism and the decay of topological defects setting $\lambda_{A}=0.1$.
		The upper line corresponds to the non-resonant production  and the lower line is the BBN bound discussed in Sec.~\ref{Resonant production}. }\label{N1prodXs}
\end{figure}

 In Fig.~\ref{N1prodXs} we show the region corresponding to the resonant sterile-neutrino production  in the $(\sin^2(2\theta),m_s)$ plane varying $X_s$. The plot includes the non-resonant production mechanism (the  upper line) as a limiting case with vanishing lepton asymmetry (see Secs.~\ref{Non-resonant production} and~\ref{Resonant production}). For each value of $X_s$ we also show the corresponding $f_a$ in two cases. The first case corresponds to a negligible $\Omega_a^{\rm string}$. In the second case we give the value of the axion decay constant, which we call $f_a^{\rm mis+string}$ there, taking into account both $\Omega_a^{\rm mis}$ and $\Omega_a^{\rm string}$ for $\lambda_A=0.1$. This is one of the values for which we can not only account for the whole DM with axions and sterile neutrinos, but we can also reproduce the observed neutrino oscillations phenomenology, baryon asymmetry, have a stable EW vacuum, critical Higgs inflation (in agreement with Planck observations~\cite{Ade:2015lrj}) and solve the strong CP problem~\cite{Salvio:2018rv}. Note that moderate variations of $\lambda_A$ around this value produce very small changes in $f_a^{\rm mis+string}$ because $\Omega_a^{\rm string}$ depends on $\lambda_A$ only logarithmically.
 
 \begin{figure}[p!]
\begin{center}
 \includegraphics[scale=0.8]{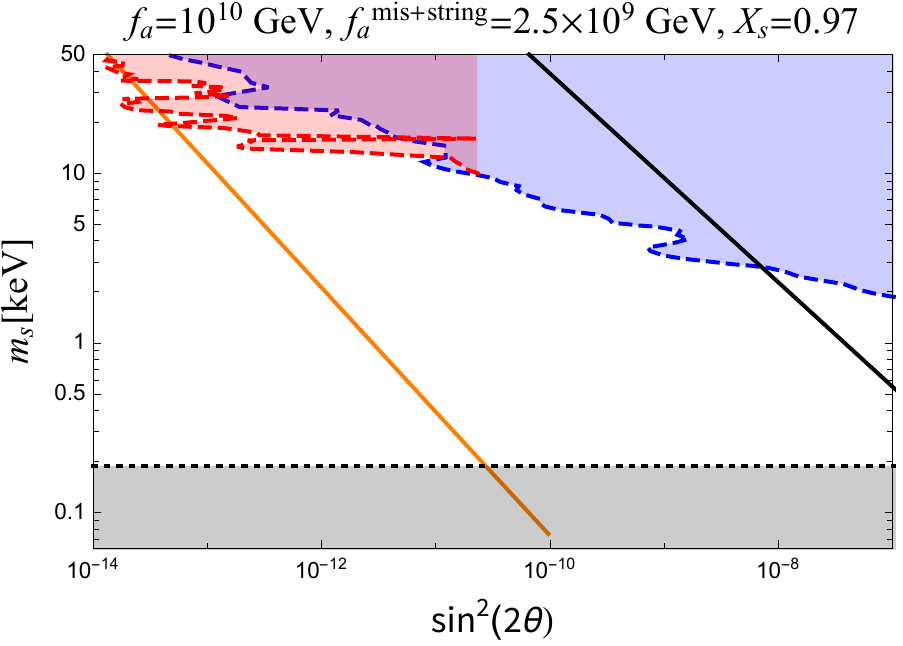}  \hspace{2cm} 
  \includegraphics[scale=0.8]{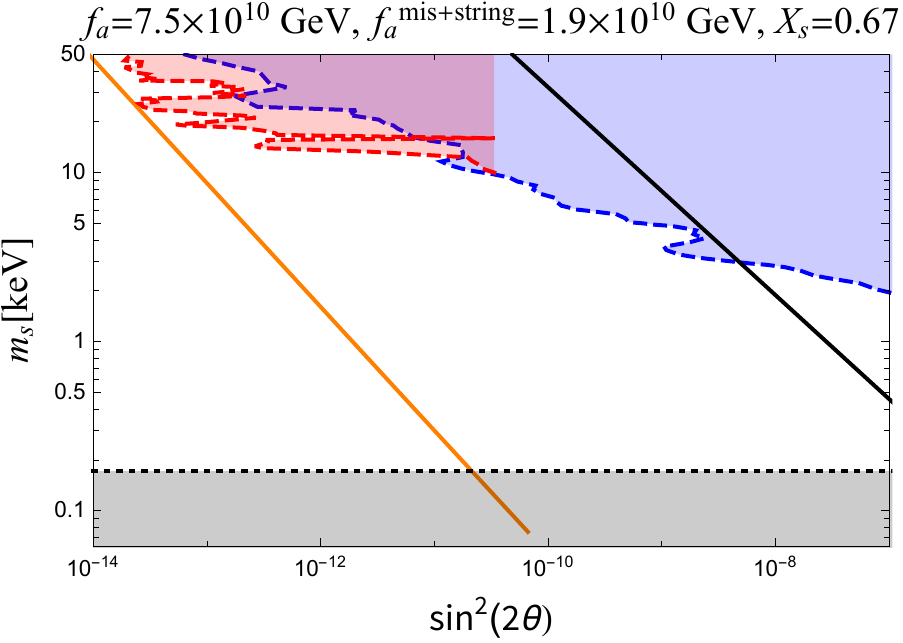} \\
  \vspace{1cm}
  \includegraphics[scale=0.8]{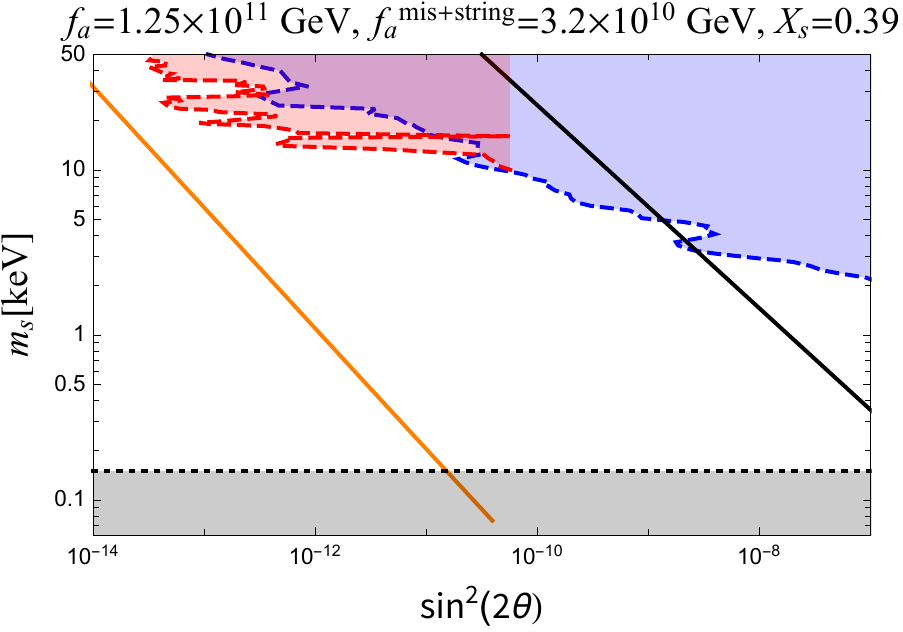}  \hspace{2cm} 
  \includegraphics[scale=0.8]{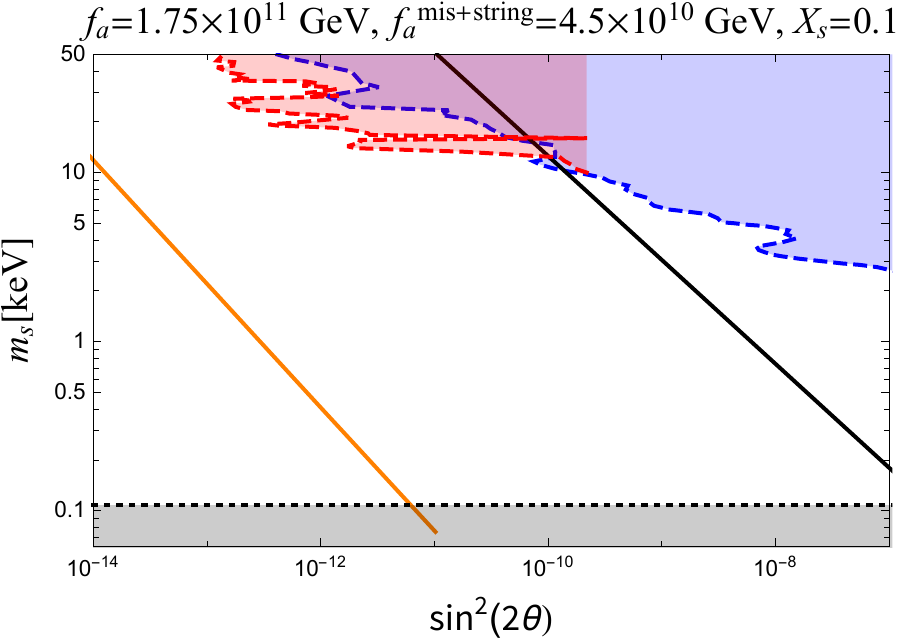} \\\vspace{1cm}
  \includegraphics[scale=0.8]{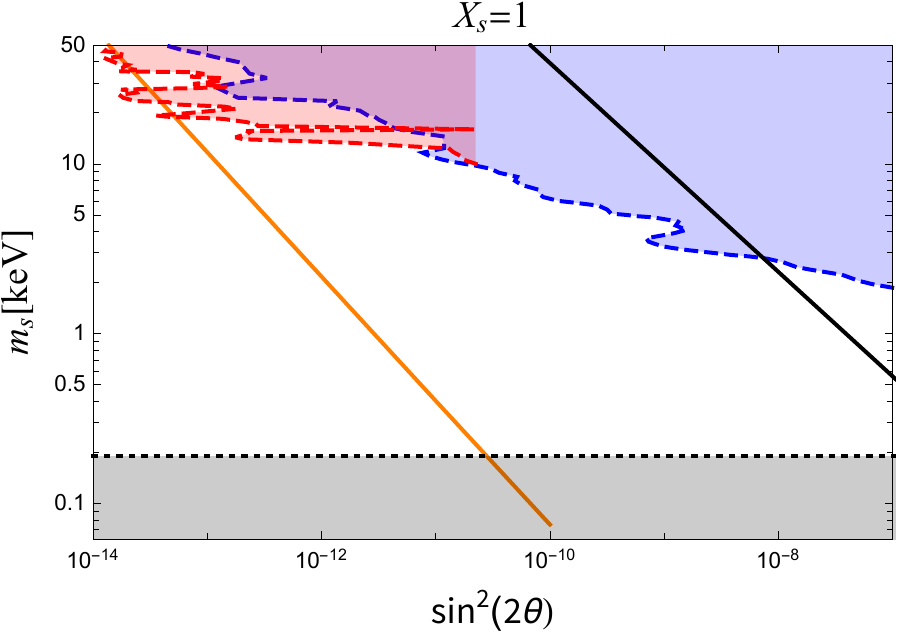}
 \end{center}
   \caption{\em 
   The upper and lower lines of Fig.~\ref{N1prodXs} (here depicted in solid black and orange, respectively) compared with the X-ray and phase-space bounds discussed in Sec.~\ref{Sterile-neutrino-DM} (dashed lines). The X-ray bounds are the upper ones in blue~\cite {Boyarsky:2018tvu} and red~\cite{Ng:2019gch},  while the phase-space ones are the lower ones in black.  In this figure we also provide the corresponding plot for $X_s=1$ (the  bottom one, see Ref.~\cite{Boyarsky:2018tvu} for a review).}
\label{N1prodXsCon}
\end{figure}

 \begin{figure}[t]
\begin{center}
 \includegraphics[scale=0.8]{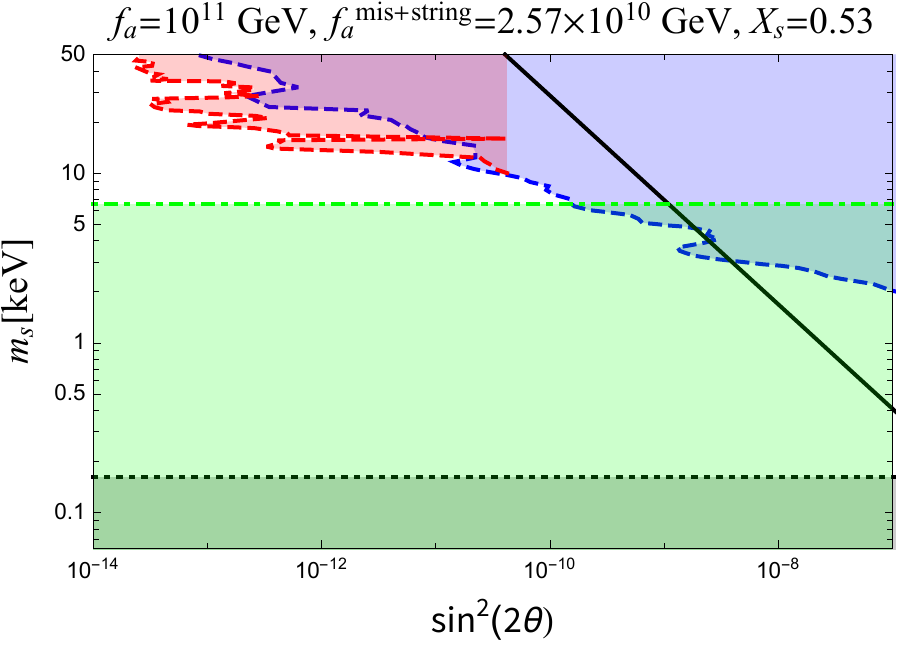}  \hspace{2cm} 
  \includegraphics[scale=0.8]{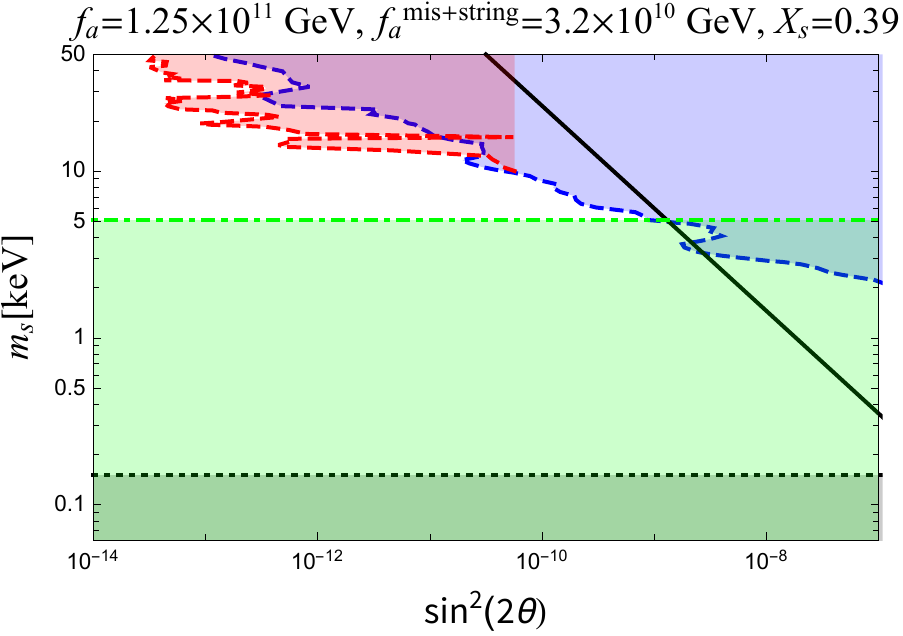} \\
  \vspace{1cm}
  \includegraphics[scale=0.8]{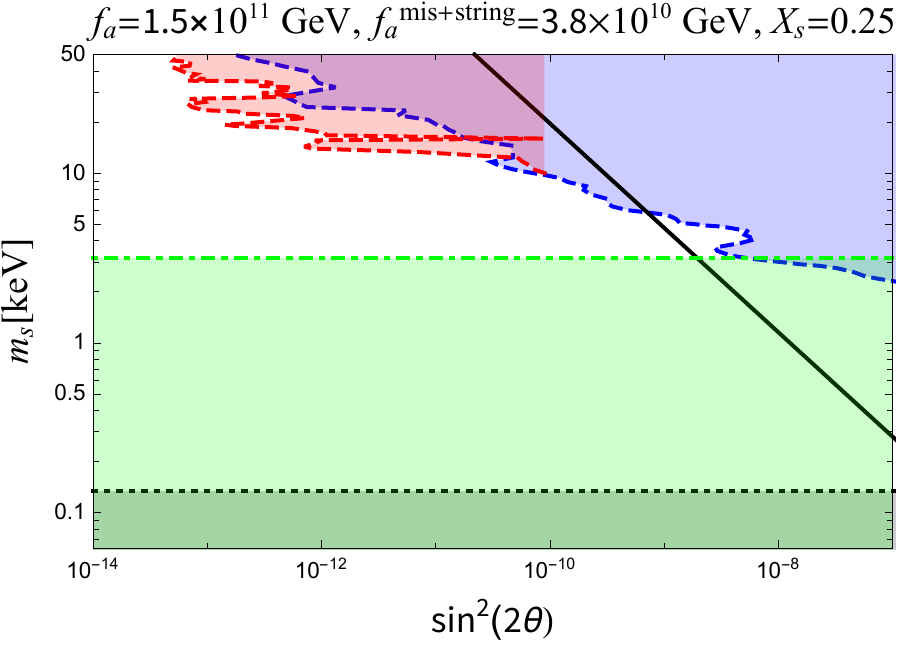}  \hspace{2cm} 
  \includegraphics[scale=0.8]{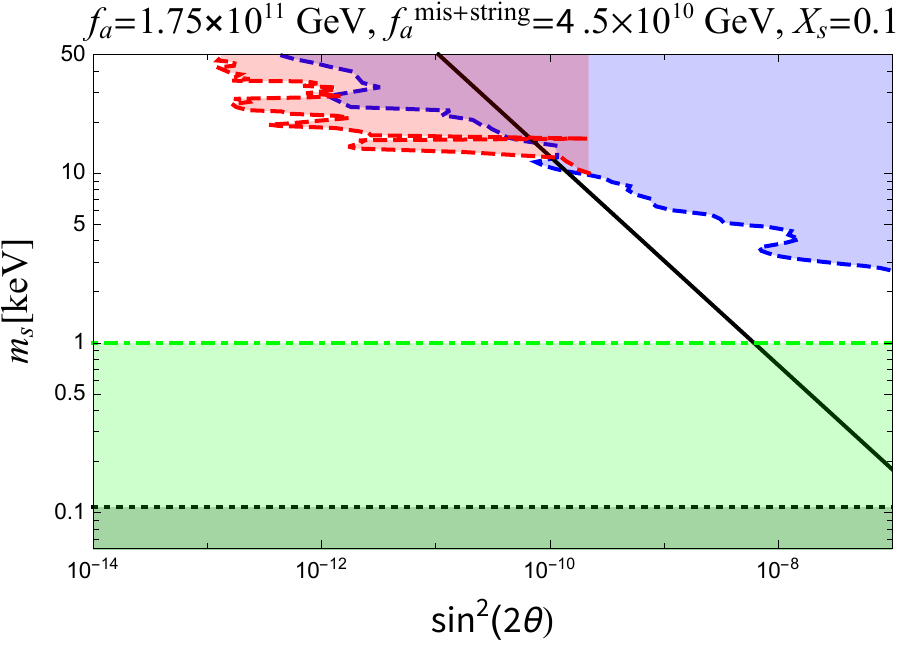} \\
 \end{center}
   \caption{\em 
   The non-resonant sterile neutrino black solid lines  and the X-ray and phase-space bounds of Fig.~\ref{N1prodXsCon} together with the structure-formation bounds of Sec.~\ref{Non-resonant production} (dash-dotted lines).}
\label{N1prodXsCon2}
\end{figure}


 In Fig.~\ref{N1prodXsCon} the  sterile-neutrino production region of Fig.~\ref{N1prodXs} is compared with the X-ray and the phase-space bounds discussed in Sec.~\ref{Sterile-neutrino-DM}. In Fig.~\ref{N1prodXsCon2} we also add the structure-formation bounds discussed in Sec.~\ref{Non-resonant production}. As shown in Fig.~\ref{N1prodXsCon2} an allowed region for non-resonant sterile-neutrino production only appears for $X_s \lesssim 0.3$.
 
Finally, regarding the CPT-symmetric universe discussed in Sec.~\ref{A mention of the CPT symmetric case}, note that, interestingly, we can express $m_s$ as a function of $f_a$ with a (mild logarithmic) dependence on $\lambda_A$.  This can be obtained by plugging Eq.~(\ref{Xsa}) into~(\ref{msCPT}) and using the expressions for $\Omega^{\rm mis}_{a}$ and $\Omega^{\rm string}_{a}$ given in Sec.~\ref{Axion-DM}.

\section{Conclusions}\label{Conclusions}

We have analysed all DM candidates in the $a\nu$MSM, a simple extension of the SM, originally proposed in~\cite{Salvio:2015cja}, which features three sterile neutrinos and the extra fields of the KSVZ QCD axion model. The $a\nu$MSM is well-motivated because it not only accounts  for DM, neutrino oscillations and baryon asymmetry, but it also solves the strong CP problem, stabilizes the EW vacuum and can implement CHI (in agreement with the most recent Planck observations).

We have ruled out PBHs as a  possible source of DM in this model because $P_\mathcal{R}$ has a peak that is several orders of magnitude below the required height. Consequently, DM in this model is generically due to the axion and the lightest sterile neutrino. Imposing several constraints, this result allows us to relate the axion parameters such as $f_a$ and $\lambda_A$ to the neutrino parameters ($m_s$ and $\theta$). 

Requiring the lightest sterile neutrino to contribute to DM in addition to the axion (the only candidate previously considered in the $a\nu$MSM) has several advantages. We have discussed how this requirement generically enlarges the parameter space with absolute EW stability and, as a result, that where CHI occurs. This inflationary scenario does not suffer from a too low scale of perturbative unitarity breaking and fine-tuning of initial  conditions (before inflation). On the other hand, the sterile-neutrino DM scenario benefits from the presence of an axion DM component because requiring the lightest sterile neutrino to account only for a fraction $X_s<1$ of the DM abundance relaxes all the existing constraints on this scenario.
Therefore, one can say that axion and sterile neutrino DM mutually reinforce each other in the $a\nu$MSM.

We plan to keep testing the $a\nu$MSM with future astrophysical and, in particular, cosmological data, such as those regarding the cosmic microwave background and structure formation.

\subsection*{Acknowledgments}
I thank G.~Ballesteros and A.~Urbano for useful discussions and J.~Rubio for useful mail communications on primordial black holes in related models.

\appendix
\section{Renormalization-group equations}
\label{RGEs}
For a generic coupling $g$ defined in the $\overline{\rm MS}$ renormalization scheme we write the RGEs as
\be \frac{dg}{d\tau}= \beta_{g},\ee
where $d/d\tau\equiv \bar{\mu}^2\, d/d\bar{\mu}^2$ and $\bar{\mu}$ is the $\overline{\rm MS}$ renormalization energy scale. The $\beta$-functions  $\beta_{g}$ can also be expanded in loops: 
\be  \beta_{g} =  \frac{\beta_{g}^{(1)}}{(4\pi)^2}+ \frac{\beta_{g}^{(2)}}{(4\pi)^4}+ ... \, ,\ee 
where $ \beta_{g}^{(n)}/(4\pi)^{2n}$   is the $n$-loop contribution. 

We start from energies much above $M_A$, $M_q$ and $M_{ij}$. In this case, the 1-loop RGEs of all relevant couplings  are~\cite{Salvio:2018rv}
%
{\allowdisplaybreaks\bea  \beta_{g_1^2}^{(1)}& =&    \frac{41g_1^4}{10}, \qquad   \beta_{g_2^2}^{(1)} =- \frac{19g_2^4}{6},\qquad\beta_{g_3^2}^{(1)}  = -\frac{19 g_3^4}{3},\nonumber\\   \beta_{y_t^2}^{(1)}  & =& y_t^2\left(\frac92 y_t^2-8g_3^2-\frac{9g_2^2}{4}-\frac{17g_1^2}{20} + {\rm Tr}(Y^\dagger Y )\right),\nonumber\\ 
  \beta_{\lambda_H}^{(1)} & =&\left(12\lambda_H+6y_t^2-\frac{9g_1^2}{10}-\frac{9g_2^2}{2}+2\, {\rm Tr}(Y^\dagger Y)\right)\lambda_H \nonumber\\ &&\hspace{-0.7cm}-\, 3y_t^4 +\frac{9 g_2^4}{16}+\frac{27 g_1^4}{400}+\frac{9 g_2^2 g_1^2}{40}+\frac{\lambda_{HA}^2}{2} - {\rm Tr}((Y^\dagger Y)^2), \nonumber\\ 
 \beta_{\lambda_{HA}}^{(1)} & =& \left(3y_t^2-\frac{9g_1^2}{20}-\frac{9g_2^2}{4}+6\lambda_H \right) \lambda_{HA}\nonumber \\ && +\left(4\lambda_A +\, {\rm Tr}(Y^\dagger Y ) + 3y^2 \right) \lambda_{HA}+2 \lambda_{HA}^2, \nonumber\\ 
 \beta_{\lambda_A}^{(1)} & =& \lambda_{HA}^2+10\lambda_A^2+6y^2 \lambda_A- 3 y^4,\nonumber\\ 
   \beta_{Y}^{(1)} & =&Y  \left[\frac32 y_t^2-\frac{9}{40} g_1^2-\frac98 g_2^2+\frac34 Y^\dagger Y+\frac12 {\rm Tr}(Y^\dagger Y )\right],\nonumber \\ 
 \beta_{y^2}^{(1)} & =&y^2(4y^2-8 g_3^2),\nonumber \\ 
  \beta_{\xi_H}^{(1)} &=& (1+6\xi_H)\left(\frac{y_t^2}{2}+\frac{{\rm Tr}(Y^\dagger Y )}6 -\frac{3g_2^2}8  - \frac{3 g_1^2}{40}+\lambda_H\right)\nonumber -\frac{\lambda_{HA}}{6}(1+6\xi_A), \nonumber \\
 \beta_{\xi_A}^{(1)}  &=& (1+6\xi_A) \left(\frac{y^2}{2}+\frac{2}{3}\lambda_A\right)
-\frac{\lambda_{HA}}{3} (1+6\xi_H), \nonumber
\eea}
where  $g_3$,  $g_2$ and  $g_1=\sqrt{5/3}g_Y$ are the gauge couplings of SM gauge group ${\rm SU(3)}_c $, ${\rm SU(2)_{\it L}}$ and   ${\rm U(1)_{\it Y}}$, respectively, $y_t$ is the top Yukawa coupling and $\lambda_H$ is the Higgs quartic coupling appearing in the term $\lambda_H(|H|^2-v^2)^2$ of  the classical potential.  

Since the SM couplings evolve in the full range from the EW to the Planck scale it is appropriate to use for them the 2-loop RGEs\footnote{In the absence of gravity the RGEs for a generic quantum field theory  were computed up to 2-loop order in~\cite{MV}.}, which, including the new physics contribution, read:
{\allowdisplaybreaks\bea 
 \beta_{g_1^2}^{(2)}&=&g_1^4 \left(\frac{199 g_1^2}{50}+\frac{27 g_2^2}{10}+\frac{44 g_3^2}{5}-\frac{17 y_t^2}{10}-\frac{3 }{10} {\rm Tr}(Y^\dagger Y)\right), \nonumber 
\\
 \beta_{g_2^2}^{(2)}&=&g_2^4 \left(\frac{9 g_1^2 }{10}+\frac{35 g_2^2}{6}+12  g_3^2-\frac{3  y_t^2}{2}-\frac{1}{2}  {\rm Tr}(Y^\dagger Y)\right),\nonumber 
\\
 \beta_{g_3^2}^{(2)}&=&g_3^4 \left(\frac{11 g_1^2}{10}+\frac{9g_2^2}{2}-\frac{40 g_3^2}{3}-2 y_t^2- y^2\right),\nonumber  \\
 \beta_{y_t^2}^{(2)}&=&+y_t^2 \bigg[ 6\lambda_H^2 -\frac{23 g_2^4}{4}+ 
y_t^2 \left( -12 y_t^2 -12 \lambda_H 
+36 g_3^2+\frac{225 g_2^2}{16}+\frac{393 g_1^2}{80} -\frac94  {\rm Tr}(Y^\dagger Y) \right)
\nonumber \\ && \hspace{-1cm}
+ \frac{1187 g_1^4}{600}+9 g_3^2 g_2^2+ 
   \frac{19}{15} g_3^2 g_1^2-\frac{9}{20} g_2^2 g_1^2  -\frac{932 g_3^4}{9}\nonumber \\ && \hspace{-1cm}
    +\left(\frac{3 g_1^2}{8}+\frac{15 g_2^2}{8} \right){\rm Tr}(Y^\dagger Y) -\frac94{\rm Tr}((Y^\dagger Y)^2)  +\frac{\lambda_{HA}^2}{2} \bigg],    \nonumber
   \\
   \beta_{\lambda_H}^{(2)}&=&\lambda_H^2 \left[54 \left(g_2^2+\frac{g_1^2}{5}\right) \right. -156\lambda_H -72 y_t^2 -24{\rm Tr}(Y^\dagger Y)\bigg] 
\nonumber \\ && \hspace{-1cm}
+\lambda_H y_t^2 \left( 40 g_3^2
+\frac{45 g_2^2}{4}+\frac{17 g_1^2}{4}-\frac32 y_t^2\right)
\nonumber \\ &&\hspace{-1cm}  +\lambda_H \bigg[\frac{1887 g_1^4}{400} -\frac{73 g_2^4}{16}
+\frac{117 g_2^2 g_1^2}{40}+\left(\frac{3 g_1^2}{4}+\frac{15 g_2^2}{4} \right){\rm Tr}(Y^\dagger Y)-\frac{{\rm Tr}((Y^\dagger Y)^2)}{2} -5 \lambda_{HA}^2\bigg]
\nonumber \\ &&\hspace{-1cm}  +y_t^4 \left( 15 y_t^2-16g_3^2-\frac{4 g_1^2}{5}\right)
 +y_t^2 \left(\frac{63 g_2^2g_1^2}{20} -\frac{9 g_2^4}{8}-\frac{171 g_1^4}{200}\right)
\nonumber \\ &&\hspace{-1cm}  +\frac{305 g_2^6}{32} -\frac{3411 g_1^6}{4000} -\frac{289 g_2^4 g_1^2}{160} -\frac{1677 g_2^2 g_1^4}{800} 
 \nonumber \\ &&\hspace{-1cm}  - \left(\frac{9 g_1^4}{200} + \frac{3 g_1^2 g_2^2}{20}+ \frac{3 g_2^4}{8}\right){\rm Tr}(Y^\dagger Y)+5{\rm Tr}((Y^\dagger Y)^3)
  -3y^2\lambda_{HA}^2-2\lambda_{HA}^3.  \nonumber \\
 \nonumber
\eea}
Here we have corrected a missprint of the RGEs provided in Ref~\cite{Salvio:2018rv}: there are no $g_3^2y^2$ and $y^4$ contributions to the RGE of $\lambda_H$.

The matching at the mass thresholds due to the new scalar $A$ and fermions $N_i$, $q_1$ and $q_2$ is performed as explained in Ref.~\cite{Salvio:2018rv}.

\vspace{1cm}

 \footnotesize
\begin{multicols}{2}

\end{multicols}

\end{document}